\documentclass[epj,final]{svjour}

\usepackage{graphics}% Include figure files

\begin{document}
% Include the paper's title here

\title{Detection of node group membership in
  networks with group overlap}

\author{Erin N. Sawardecker\inst{1} \and Marta Sales-Pardo\inst{1,2,3}
\and Lu\'{\i}s A. Nunes Amaral\inst{1,2}}

\mail{amaral@northwestern.edu}

\institute{Department of Chemical and Biological Engineering,
Northwestern University, Evanston, IL 60208, USA \and Northwestern
Institute on Complex Systems, Northwestern University, Evanston, IL
60208, USA \and Northwestern University Clinical and Translational
Sciences Institute, Chicago, IL 60611, USA}

\date{Received: 1 August 2008}

\abstract{ Most networks found in social and biochemical systems have
modular structures. An important question prompted by the modularity
of these networks is whether nodes can be said to belong to a single
group. If they cannot, we would need to consider the role of
``overlapping communities.'' Despite some efforts in this direction,
the problem of detecting overlapping groups remains unsolved because
there is neither a formal definition of overlapping community, nor an
ensemble of networks with which to test the performance of group
detection algorithms when nodes can belong to more than one
group. Here, we introduce an ensemble of networks with overlapping
groups. We then apply three group identification methods---modularity
maximization, {\it k}-clique percolation, and modularity-landscape
surveying---to these networks. We find that the modularity-landscape
surveying method is the only one able to detect heterogeneities in
node memberships, and that those heterogeneities are only detectable
when the overlap is small. Surprisingly, we find that the {\it
k}-clique percolation method is unable to detect node membership for
the overlapping case.   }

\maketitle
%\tableofcontents
%%%%%%%%%%%%%%%%%%% END OF PREAMBLE %%%%%%%%%%%%%%%%%%%%%%%%

%%%%%%%%%%%%%%%%%%%%%%%%%%%%%%%%%%%%%%%%%%%%%%%%%%%%%%%%%%%%
%%%%%%%%%%%%%%%%%%% BODY OF THE PAPER %%%%%%%%%%%%%%%%%%%%%%
%%%%%%%%%%%%%%%%%%%%%%%%%%%%%%%%%%%%%%%%%%%%%%%%%%%%%%%%%%%%

\section{Introduction}
\label{sec:intro}

% ------------------ Introduction -----------------------
Real-world networks including man-made and natural networks are
strongly modular, that is, the pattern of connections among nodes is
not homogeneous~\cite{newman04b,guimera05}. The modularity of a
network is a consequence of the fact that there are groups of nodes in
the network that preferentially connect to one another
\cite{newman04b,guimera05,watts02,donetti04,guimera04,reichardt04,duch05,palla05}.
However, the assignment of nodes into those groups still remains a
challenging task because, typically, nodes also connect to nodes that
are not in their group \cite{watts02,reichardt04,palla05}.
Additionally, nodes may hold membership in more than one group
\cite{reichardt04,palla05,baumes05,gfeller05,nepusz08}, resulting in
groups that ``overlap.''

The question of whether there are nodes that belong to more than one
group and how important overlapping groups are to the network's
organization is especially relevant in social and biochemical systems,
in which typically nodes are thought to belong to more than one
group. Consider, for instance, the network of scientific
collaborations within an institution: People with joint appointments
would be expected to appear in more than one group. Or, consider the
network of physical interactions between proteins: Topological modules
are thought to have a strong correlation with biological
function~\cite{pereira-leal04}. Since many proteins are known to have
more than one function, one would expect these proteins to belong to
more than one group.

Recently, methods to uncover the hierarchical organization of
networks~\cite{sales-pardo07,clauset08} have been proposed,
opening the possibility of performing multi-scale analysis on social,
biological, and economical systems for which large amounts of data are
available. However, a potential caveat of these methods is that they
do not take into account the fact that nodes could hold membership in
more than one group. Modularity maximization
methods~\cite{duch05,guimera05a,newman06}, which have been successful
at finding correlations between network function and structure, suffer
from the same problem. {\it The impact of neglecting overlapping groups has
not been assessed so far, since there is neither a formal
definition of overlapping group nor a set of models on which to
test overlap identification algorithms.}

Here, we introduce an ensemble of networks~\cite{newman03h} that have
overlapping groups by construction~\cite{nepusz08}. We then apply
three different group detection methods---modularity
maximization~\cite{newman04b,guimera05a}, {\it k}-clique
percolation~\cite{palla05}, and modularity-landscape
surveying~\cite{sales-pardo07}---to these networks. We find that the
modularity-landscape surveying method is the only one able to detect
heterogeneities in node memberships, and that these heterogeneities
are detectable provided the overlap is significantly smaller than the
size of the modules involved.

%%%%%%%%%%%%%%%%%%%%%%%%%%%%%%%%%%%%%%%%%%%%%%%%%
%Description of the models
%%%%%%%%%%%%%%%%%%%%%%%%%%%%%%%%%%%%%%%%%%%%%%%%%%
\section{Model networks}
\label{sec:model}

Consider a network comprised of $N$ nodes and $L$ edges. Let
$\mathcal{N} =$ \{$\:n_i: i = 1, \dots, N\:$\} be the set of nodes and
$\mathcal{G}=\{\:g_x: x= a, b, \dots \:\}$ be the set of groups in
which the nodes can hold membership. Specifically, let
$\mathcal{G}_i\subset \mathcal{G}$ be the set of groups in which node
$n_i$ holds membership. Without loss of generality, we assume that all
groups with identical membership lists have been merged and that all
groups have at least two members.

Here, we focus on the ensemble of random networks in which
the probability $p_{ij}$ of the edge $(n_i, n_j)$ being present in a
network is a function {\it solely\/} of the set of co-memberships of 
the two nodes $\mathcal{M}_{ij} = \mathcal{G}_{i} \cap \mathcal{G}_{j}$. 
We restrict our attention to the case where $p_{ij}$ is a
non-decreasing function of the cardinality of $\mathcal{M}_{ij}$. For
undirected networks, this is the most plausible case. Indeed, most
clustering algorithms used to investigate the modular structure of
networks have at their basis the assumption that this is the
appropriate case. Such an assumption is supported by the fact that
those methods return plausible results for those networks
\cite{duch05,newman03h,guimera05b,hastings06,newman06b}.

We consider the case where $p_{ij} = p_l$, where
$l=||\mathcal{M}_{ij}||$ is the cardinality of $\mathcal{M}_{ij}$,
and, further, we assume that $p_0 < p_1 \leq p_2 \leq \cdots$. An
implication of our choice for $p_{ij}$ is that if one selects a
sub-set of $\mathcal{N}$ in which all pairs of nodes have non-empty
co-membership sets, then there will be more edges connecting these
nodes than one would expect to find by chance. Or, in other words, we
expect to find more edges connecting these nodes than if all links had
the same probability $p = 2L / N(N-1)$ of being present. In contrast,
if one selects a sub-set of $\mathcal{N}$ in which all pairs of nodes
have empty co-membership sets, then there will be fewer edges
connecting these nodes than one would expect to find by chance. These
facts directly suggest that the maximization of a modularity function
such as that proposed by Newman and Girvan \cite{newman04b} will
enable one to identify node membership in modular networks (see
\cite{sales-pardo07,fortunato07,kumpula07} for caveats to this
argument).
 
The ensemble of networks we focus on comprises two distinct
sub-ensembles. The first sub-ensemble, which we denote {\it
transitive\/}, conforms to a transitive relationship among
co-membership sets. That is, if the co-membership set $\mathcal{M}_{ij}$
is non-empty and the co-membership set $\mathcal{M}_{ik}$ is also
non-empty, then the co-membership set $\mathcal{M}_{jk}$ must also be
non-empty. The second sub-ensemble, which we denote {\it
non-transitive\/}, does not conform to a transitive relationship
among co-membership sets.
 
Networks in the transitive sub-ensemble have the property that every
node must hold membership in only one group (if identical groups have
been collapsed). In contrast, networks in the non-transitive
sub-ensemble have some nodes that hold membership in more than one
group. Most module detection algorithms in the literature deal only
with the transitive sub-ensemble
\cite{duch05,guimera05a,newman06b,danon05}, that is, nodes are divided
into ``exclusive groups.'' A notable exception is the work of Palla et
al.  \cite{palla05}, which highlights the possibility that a network
will contain nodes belonging to more than one group, thus allowing for
``overlapping communities.'' Regretfully, Palla et al. \cite{palla05}
do not define ensembles of networks with overlapping groups. 

\begin{figure*}
\resizebox{2\columnwidth}{!}{%
  \includegraphics{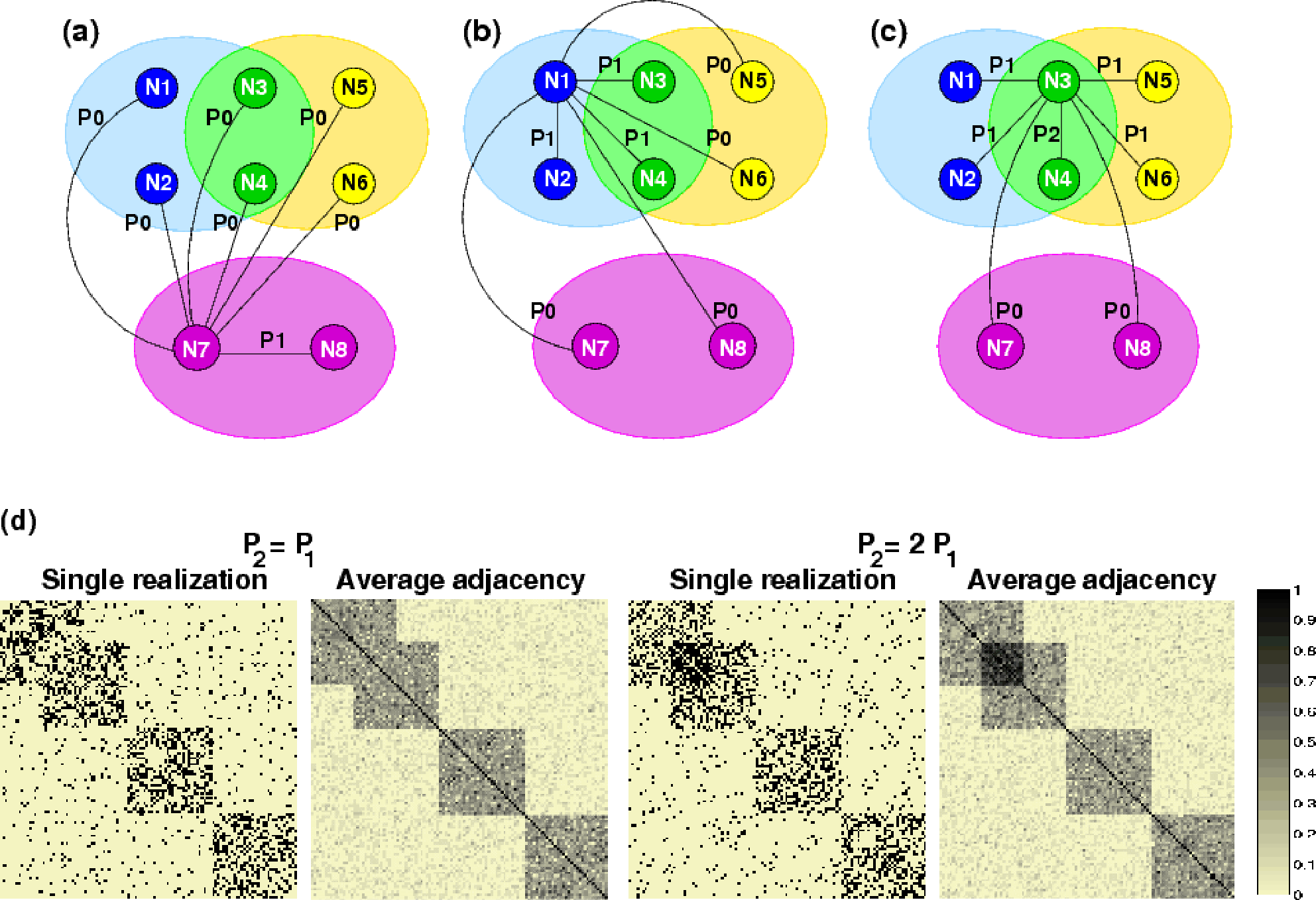}
}
\caption{ (Color online) Random network models with overlapping
groups.  Consider a network with eight nodes, $n_1$ through $n_8$. We
assign these nodes to three different groups which we indicate by
three different colors, blue, yellow, and purple. Moreover, we allow
the blue and yellow groups to overlap; nodes $n_3$ and $n_4$ belong to
both yellow and blue groups. We highlight these overlapping nodes in
green. ({\bf a}), Purple nodes connect to other purple nodes with
probability $p_1$ and to blue, green, and yellow nodes with
probability $p_0 < p_1$. ({\bf b}), Blue nodes connect with other
nodes in the blue group ($n_1$, $n_2$, $n_3$, $n_4$) with probability
$p_1$, and to nodes outside their group (yellow and purple nodes) with
probability $p_0$. Note that yellow nodes have the same properties as
blue nodes. ({\bf c}), Green nodes connect to blue and yellow nodes
with the same probability $p_1$. Since green nodes are not in the same
group as purple nodes, the probability of having a connection between
a green and purple node is $p_0$. For generality, we assume that green
nodes connect between themselves with probability $p_2$. ({\bf d}),
Adjacency matrices for the model networks with $p_2 = p_1$ and $p_2 =
2 p_1$. We show both the adjacency matrix for a single realization of
the model network and the average adjacency matrix, that is the
fraction of times two nodes are connected in the ensemble of networks
defined by the model. Note that $p_2 = 2 p_1$ has an overlapping
region that is more densely connected, hence the average adjacency
matrix has more ``black.'' We show results for networks of 112 nodes
divided into four groups of 32 nodes. Two of the four groups overlap
by sharing 16 nodes and the corresponding edges. We select $p_1$ and
$p_0$ such that for nodes not in the overlapping region, the average
degree $z$ is 16 and the ratio $r$ between the external-degree and
internal-degree is $r = 0.125$. }
\label{f.overlap_networks_schematic1}
\end{figure*}

Another significant exception is the work of Sales-Pardo et
al. \cite{sales-pardo07}, which determines community structure even
when hierarchical levels of structure exist. These hierarchical levels
of structure indicate that nodes may belong to more than one group,
but only when multiple layers are considered. Here, we define an
ensemble of networks in which most nodes hold membership in a single
group, while a small fraction of nodes hold membership in two or three
groups (Fig.~\ref{f.overlap_networks_schematic1}).

%%%%%%%%%%%%%%%%%%%%%%%%%%%%%%%%%%%%%%%%%%%%%%%%%
%Description of the methods
%%%%%%%%%%%%%%%%%%%%%%%%%%%%%%%%%%%%%%%%%%%%%%%%%%

\section{Community detection}
\label{sec:methods}

\subsection{Description of the methods}
\label{sec:detection}

Let us now address the question of detectability of the memberships of
individual nodes. Ideally, one wishes to detect all group
memberships from the topology of the network alone. For the case of
transitive networks, it has already been shown that when $p_1$, the
probability that two nodes belonging to the same group are connected,
is not much larger than $p_0$, it is impossible to extract the correct
membership assignment from the network structure alone
\cite{guimera05a}. Here, we focus on the detection of node membership
for the ensemble of non-transitive networks described above. We
consider three different classes of group detection algorithms:
modularity maximization \cite{newman04b,guimera05a}, {\it
k}-clique percolation \cite{palla05}, and
modularity-landscape surveying \cite{sales-pardo07}.

Modularity maximization methods are the current ``gold standard'' for
group identification \cite{danon05}. In this approach, nodes are
classified into groups that maximize the number of within group edges
compared to those that would be expected from chance
alone~\cite{newman04b,guimera05,donetti04,guimera04,duch05}. Some of
the proposed algorithms, such as spectral decomposition, are extremely
fast and can handle networks comprised of hundreds of thousands of
nodes \cite{duch05}.  However, this approach is clearly geared toward
networks with transitive membership structures since every node must
be classified into a single group.

The {\it k}-clique percolation method introduced by Palla et
al.~\cite{palla05} is based on the observation that networks sometimes 
contain connected cliques of the same size \cite{palla05}. In this
method, a group comprises chains of ``adjacent'' {\it k}-cliques---where 
two {\it k}-cliques are adjacent if they share $k-1$ nodes. 
A strength of this approach is that nodes can be classified into more
than one group, making it {\it a priori} well-suited to investigate 
non-transitive networks. Two limitations of this this approach, 
however, are that different values of $k$ will result in different 
group membership patterns, and that sparse networks might contain 
a very small number of cliques with $k>2$.

The modularity-landscape surveying method~\cite{sales-pardo07}, or
MLS, is based upon the observation that the modularity landscape is
very rugged and has many local maxima, which means that there are many
partitions of nodes into groups characterized by high values of the
modularity function. In analogy to disordered physical systems whose
landscapes are also rugged
~\cite{stillinger82,stillinger84,stillinger85}, one expects that the
set of all local maxima conveys the relevant contribution to the
system's physical properties. Thus, the method samples all the
partitions $\mathcal{P}$ corresponding to local maxima with
probabilities proportional to the size of their basins of
attraction. Then, it builds a co-classification matrix ${\rm \bf A}$,
in which each element ${\rm A}_{ij}$ corresponds to the expected
fraction of the time in which a pair of nodes $(n_i,n_j)$ is
classified in the same group. As discussed above, this method does not
restrict nodes to hold membership in a single group, but rather, by
collecting statistics, it reports the likelihood that two nodes are
members of the same group (or sets of groups), and therefore it is
{\it a priori\/} suitable to identify node memberships in networks
with both transitive and non-transitive memberships.

\subsection{Random network ensembles}
\label{sec:network_ensembles}

In order to investigate the detectability of the membership structure
of a network, we generate random networks in which nodes can belong to
five groups, ${\cal G}\:= \{\:g_a,\:g_b,\:g_c,\:g_d,\:g_e\:\}$. We consider 
the cases in which most of the nodes belong to a single group, 
$\{g_a\}, \{g_b\}, \{g_c\}$, or $\{g_d\}$, and the remaining nodes belong 
to two groups, $\{\:g_a, g_b\:\}$, or to three groups, 
$\{\:g_a,\:g_b,\:g_e\:\}$. 
We then assume that if two nodes have membership in the same group, they 
will be linked with probability $p_1\:>\:p_0$. Similarly, if two nodes have 
membership in the same two groups, they will be connected with probability 
$p_2\:\geq\:p_1$. Note that as $ r\:\equiv\:p_0 / p_1$ approaches one,
the harder it becomes to detect the co-membership structure of a network. 
Also, since some nodes belong to multiple groups, the degree of these 
overlapping nodes will be larger than that for a node belonging to 
only one group. Henceforth, we denote the average degree of a node belonging 
to only one group by $z$. 

If a node holds membership in only one group, say $g_a$, then it
belongs to group $A$. If some nodes
hold membership in two groups \{$\:g_a,\:g_b\:$\}, then they belong to
groups $A$ and $B$. If the latter case is true, then groups $A$ and
$B$ overlap. We define the ``overlap size'' $s$ as
\begin{equation}
s \: = \: \frac{||A\cap B||}{||A \cup B||}~,
\label{e.ov-size}
\end{equation}
that is, the number of nodes in both $A$ and $B$ divided by the
combined size of groups $A$ and $B$. Thus, an important issue
regarding co-membership detection when nodes hold membership in more
than one group is how the size of the overlap affects the accuracy
in detecting group membership.

\subsection{Mutual information}
\label{sec:mutual_info}

To quantify the similarity between two partitions of nodes, we 
calculate the mutual information between the two partitions~\cite{danon05}: 

%%########### Mutual Info Equation #####################
%
\begin{equation}
\label{e.mutual_information}
M_I = \frac{-2\sum_{P,Q} N_{ij} \ln(\frac{N_{ij} N}{N_i N_j})}{\sum_{P} N_i \ln(\frac{N_i}{N}) + \sum_{Q} N_j \ln(\frac{N_j}{N})} ,
\end{equation}
%
%%######################################################
where $P$ is the list of groups in the first partition, $Q$ is the
list of groups in the second partition, $N$ is the total number of
nodes, $N_i$ is the number of nodes in group $g_i$ in the first
partition, $N_j$ is the number of nodes in $g_j$ in the second
partition, and $N_{ij}$ is the number of nodes that are both in $g_i$
and $g_j$. Note that this expression is symmetric; thus, it is an
unbiased metric to compare the similarity of two partitions.

If the partitions are identical, $M_I=1$, whereas if the two
partitions are totally uncorrelated, $M_I=0$. Note, however, that for
the case in which each node is placed into a separate group, one has
$M_I=M_I^{\ast}>0$.  We thus report
$m_I=\frac{M_I-M_I^{\ast}}{M_I^{\ast}}$, so that values of $m_I$
greater than zero indicate significant accuracy.

%%%%%%%%%%%%%%%%%%%%%%%%%%%%%%%%%%%%%%%%%%%%%%%%%
%Description of the results
%%%%%%%%%%%%%%%%%%%%%%%%%%%%%%%%%%%%%%%%%%%%%%%%%
\section{Results}
\label{sec:results}

\begin{figure*}
\resizebox{2\columnwidth}{!}{
\includegraphics{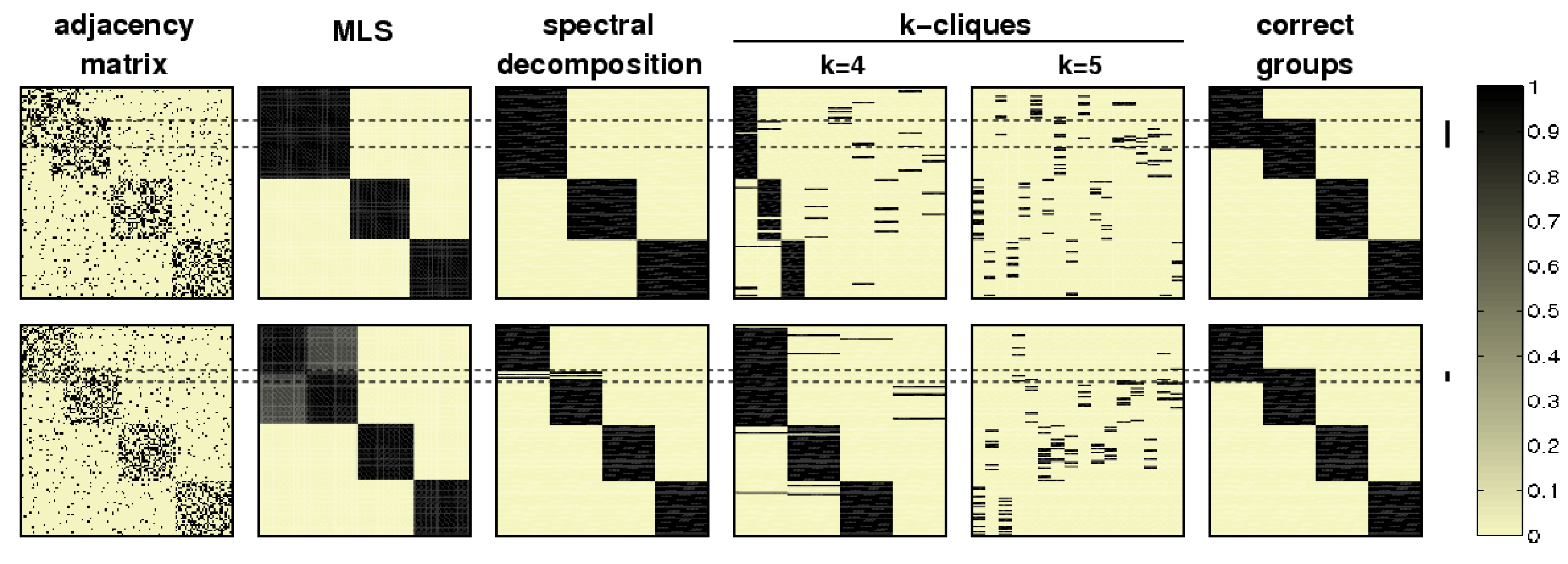}}
\caption{ (Color online) Output of methods for group detection. We apply three
different group detection algorithms, the spectral decomposition
method, the modularity-landscape surveying method, and the {\it
k}-clique method to model networks with overlapping groups
(Fig.~1). We show results obtained for a typical network realization
for $p_2 = p_1$
%in the two top rows and for $p_2 = 2 p_1$ in the two bottom rows and 
for two overlap sizes: large overlap ($s=0.25$---first
row), and small overlap ($s=0.125$---second row). We generate these
networks with the same parameters $z$ and $r$ that we use for the
networks in Fig.~1.
The first column shows the adjacency matrix. The second column shows
the co-classification matrix obtained with the modularity-landscape
surveying method (see text). The third column displays the group
classification matrix obtained with the spectral decomposition
method. The next two columns show the group classification matrices
obtained with the {\it k}-clique method for $k=4$ and $k=5$. The final
column shows the expected group classification. The colorbar is the
same as in Fig.~\ref{f.overlap_networks_schematic1}.
In the group classification matrices, each row corresponds to a node,
and each column corresponds to a group. If a node belongs to the
group, a black band appears in that group's column. If a node belongs
to more than one group, multiple bands will appear; these are nodes
that the belong to the ``overlap'' between groups according to the
algorithm. Dotted black lines indicate the nodes that by construction
hold membership in more than one group and thus comprise the
overlap in the model network.
The modularity-landscape surveying method indicates a region of
membership heterogeneity such that nodes in the overlapping groups
have more in common with each other than with nodes in the
non-overlapping groups (grey region), but this effect is dependent on
the value of $s$. The spectral decomposition method yields three or
four groups, depending on the value of $s$. The {\it k}-clique method
yields at least four groups for $k \geq 4$, some of which only contain
a few nodes.
%
%For $p_2 = 2 p_1$ with a large overlap, the $k = 5$ matrix displays a similar group pattern to the 4-clique results. Note, however, that the {\it k}-clique method incorrectly places some nodes that do not comprise the network's overlap into multiple groups, while at other times, some nodes that comprise the overlap are placed into only one group. 
}
\label{f.coclass_matrix_both_models}
\end{figure*}

To compare the performance of the methods for the ensemble of model
networks with overlapping groups previously introduced, we
generate ten networks for each set of parameter values and apply the
three group detection algorithms to each network.
Figure \ref{f.coclass_matrix_both_models} displays the typical results 
obtained for a network with parameters $r = 0.125$, $z = 16$, $p_2 = p_1$ 
or $p_2 = 2p_1$, and $s = 0.125$ or $s = 0.25$. 

\begin{figure*}
\resizebox{2\columnwidth}{!}{
\includegraphics{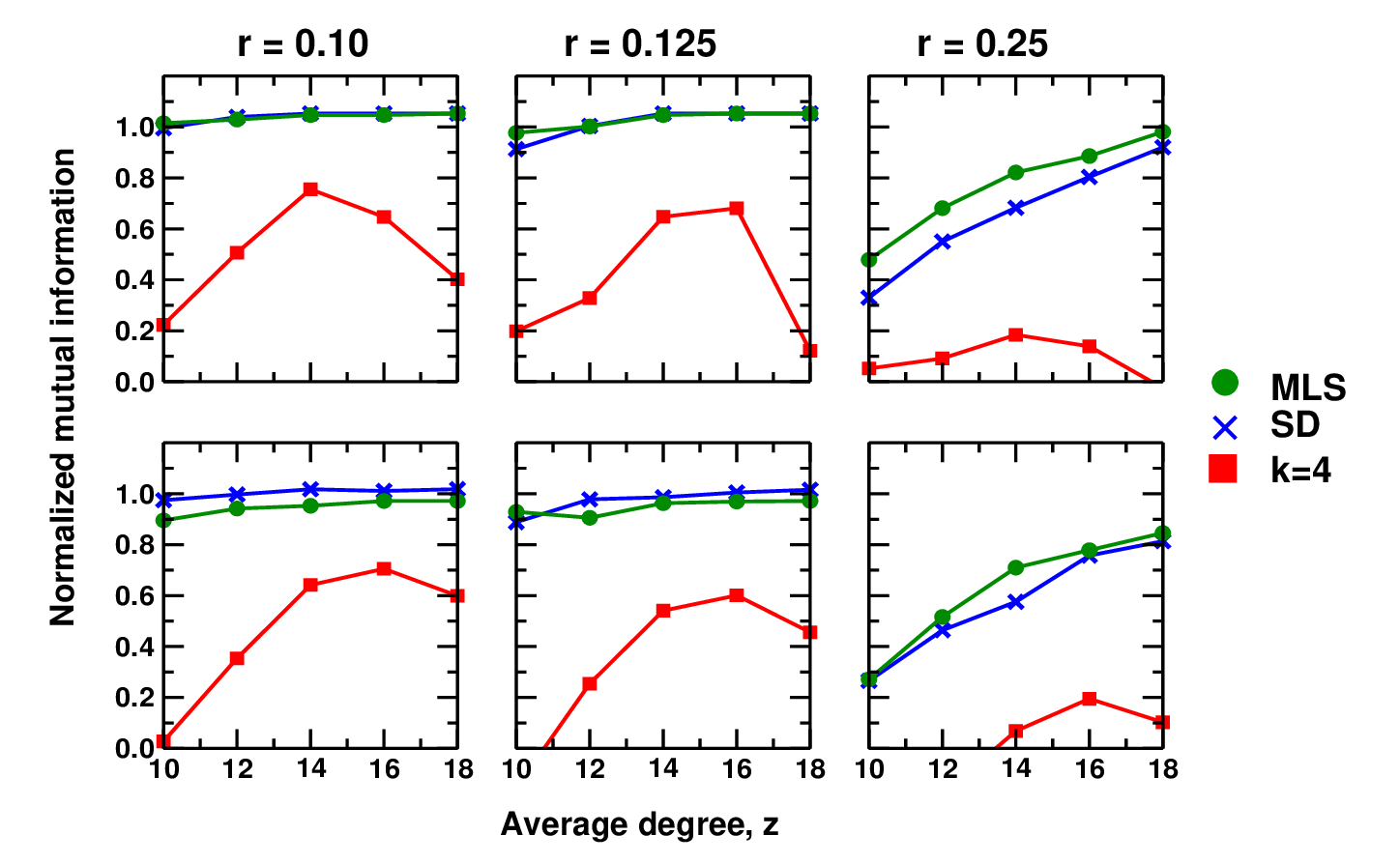}}
\caption{ (Color online) Performance of modularity and {\it k}-clique
methods for group detection. We generate ten networks for each set of
parameter values, {\it z} and {\it r}. We then compare the groups we
obtain from the {\it k}-clique, spectral decomposition, and
modularity-landscape surveying methods to the expected group
assignment and compute $m_I$ (see text).  We show the normalized
mutual information, $m_I$, versus average node degree $z$ for networks
with $p_2 = p_1$ and a large overlap (first row) and a small overlap
(second row) (Fig.~\ref{f.overlap_networks_schematic1}). For the
modularity-landscape surveying method, we computed $m_I$ with regards
to the topmost level only. The first column of plots corresponds to $r
= 0.1$, the second column to $r = 0.125$, and the third column to $r =
0.25$. Note that $m_I > 0$ indicates that the method considered yields
significant information on the network's group structure. For the {\it
k}-clique method, $m_I \leq 0$ indicates either the lack of detectable
cliques of that size or the tendency to put all nodes in the same
group. Note that,
%except for the {\it k}-clique method with $k=3$, 
for the spectral decomposition and modularity-landscape surveying
methods, performance increases with increasing average node degree $z$
and decreasing $r$. }
\label{f.mutualinfo_overlap_1pi}
\end{figure*}

To determine the accuracy of each method, we compare the partitions
returned by each method to the known division of nodes into groups.
Specifically, we use the normalized mutual information $m_I$, which
quantifies the amount of information that two different partitions
share~\cite{danon05}. Figure~\ref{f.mutualinfo_overlap_1pi} displays
the average $m_I$ versus $z$ for $p_2 = p_1$ and $s = 0.125$ or $s =
0.25$ and different values of $r$.

Since the average degree of a node should strongly affect the ability
of each group detection method to detect the known group structure, we
systematically investigate degree effects. We expect that, as degree
increases, the difficulty of detection should decrease. Also, as the
number of nodes having membership to two or more groups increases, the
difficulty of detection should increase.

\subsection{Modularity maximization}
\label{sec:results_spectral_decomp}
%The spectral decomposition method typically identifies three groups for large overlaps, and four groups for small overlaps, which is consistent with the pattern of dense interconnected regions in the network, since $m_I>0$ for any value of $z$ and $r$. 

The results obtained with the spectral decomposition method exhibit
different behaviors depending on the size of the overlap: for small
overlaps ($s = 0.125$), the method identifies four groups, whereas for
large overlaps ($s = 0.25$), it identifies three groups, such that the
two overlapping groups are combined into a single one, as predicted by
Fortunato and Barth\'{e}lemy \cite{fortunato07}. Note that there are
no significant differences between the cases $p_2=p_1$ and $p_2=2p_1$.

\subsection{Modularity-landscape surveying}
\label{sec:results_MLS}
In contrast, the modularity-landscape surveying method is able to
uncover more information about the underlying organization of the
nodes in the network than either the modularity maximization or {\it
k}-clique percolation methods. Even for small overlaps, the algorithm
is able not only to identify densely interconnected groups of nodes,
but is also able to detect that the overlapping groups have more in
common with each other than with the remaining groups
(Fig.~\ref{f.coclass_matrix_both_models}).
 
\subsection{ {\it k}-clique percolation}
\label{sec:results_k-clique}
The results obtained with the {\it k}-clique method depend strongly on
the value of $k$. For $k = 3$, the method is unable to detect the
modular structure of the networks; it places all the nodes into a
single group. For $k = 4$, the two overlapping groups are mostly
combined into one group for both large and small overlaps. Finally,
for $k = 5$, the algorithm does not identify any sizable group of
nodes in the network. In fact, the signal provided by the {\it
k}-clique method is weaker than that provided by the adjacency
matrix. Surprisingly, even though the {\it k}-clique method allows nodes
to belong to more than one group, {\it the nodes placed in multiple
groups do not in general correspond to the nodes belonging to the
overlapping groups} (Fig.~\ref{f.coclass_matrix_both_models}).

We find that the accuracy of the {\it k}-clique method is always much
smaller than that of the spectral decomposition and
modularity-landscape surveying methods. In fact, in order for the {\it
k}-clique method to return results that are significant, one must have
$r<0.25$. Moreover, for $k=4$, the accuracy of the method decreases as
the density of edges increases.

For low edge densities, the network does not contain any 5-cliques, so
the groups identified for $k=5$ are unreliable. These results point to
a severe limitation of the {\it k}-clique method: similar networks
require different {\it k} values in order to yield meaningful results,
and even when group detection is meaningful the method always performs
significantly worse than modularity based methods.

\subsection{Overlap detectability}
\label{sec:results_overlap_detectability}
\begin{figure*}
\resizebox{2\columnwidth}{!}{
\includegraphics{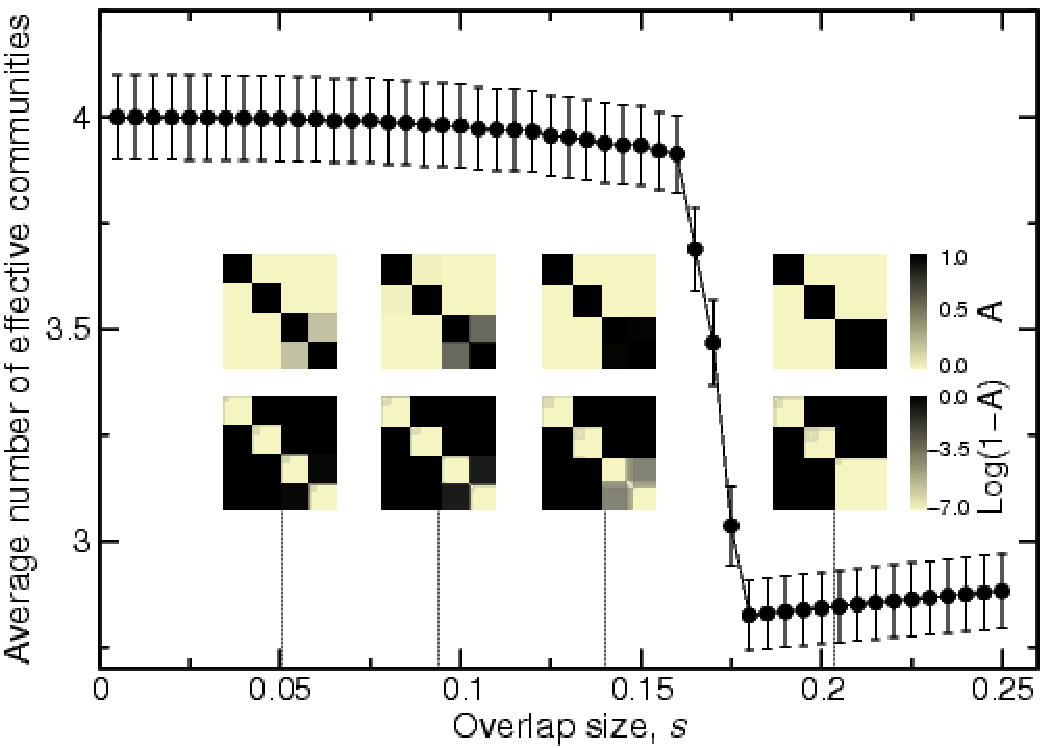}}
\caption{ (Color online) Effect of overlap size. We plot the average number of
groups detected using the spectral decomposition method for modularity
maximization (black circles) versus the overlap size $s$
(Eq. (\ref{e.ov-size})) for model networks with groups of $100$ nodes,
$z=50$, $r=0.1$, and $p_2= 2\:p_1$. We show the average and standard
error obtained for forty networks. We also show the co-classification
matrix ${\rm \bf A}$ obtained from the modularity-landscape surveying
method for a typical network for four different $s$ values. In the
co-classification matrices, each row/column represents a node and each
element ${\rm A}_{ij}$ corresponds to the fraction of the time two
nodes are classified in the same partition for the maxima in the
modularity landscape (see text). Each element is colored
following the color code on the right hand side. To illustrate the
differences for co-classification values close to one, we also plot
$\log({\rm 1 - A_{ij}})$. Again, matrix elements are colored following
the color code on the right hand side. Note that while the spectral
decomposition method can only detect whether the overlap is large
enough to transition between four to three groups, the
modularity-landscape surveying method is able to capture the greater
affinity between the nodes in two of the groups in the organization of
the network.
}
\label{f.resolution}
\end{figure*}

These results suggest that the detection of overlapping groups may be
essentially impossible when the overlap is large. However, for small
overlaps, the modularity-landscape surveying method is able to detect
heterogeneities in node group membership. The question that arises is
thus how small should the overlap be in order to be detected and
whether detection may ever be unambiguous. To answer this question, we
analyze model networks with groups comprising $100$ nodes, $z=16$,
$r=0.1$, and $p_2=2p_1$, for a wide range of overlap sizes
$s=0.02,\dots,0.25$ (Figs.~\ref{f.resolution} and
\ref{f.resolution_limits}). The spectral decomposition method shows a
transition from identifying four groups ($s<0.175$) to identifying
three groups ($s>0.2$). For $s<0.175$, the modularity-landscape
surveying method is able to detect the signature of heterogeneities in
node membership. However, the ``signal'' fades as $s$ increases.

\begin{figure*}
  \resizebox{2\columnwidth}{!}{
    \includegraphics{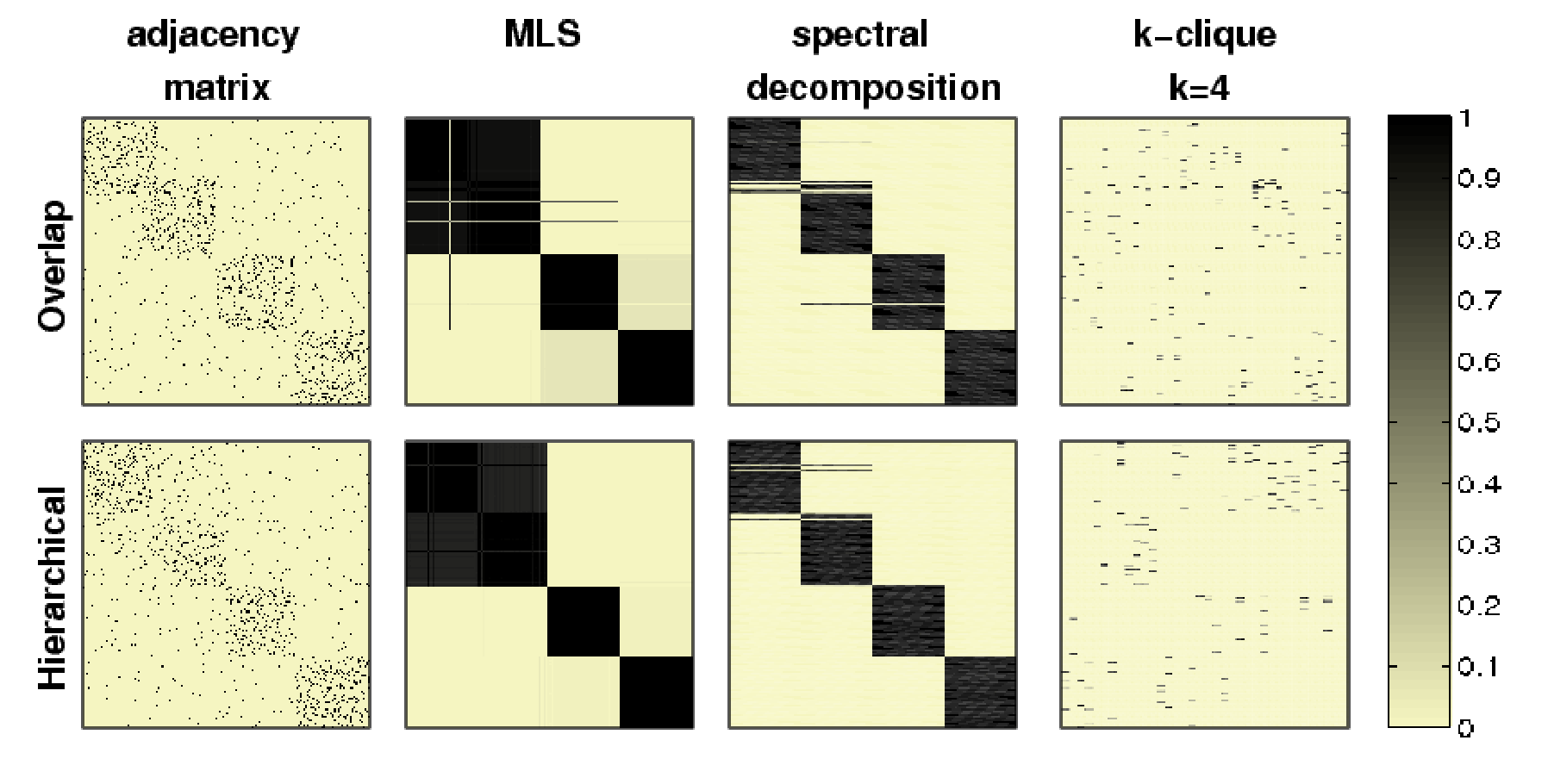}}
  \caption{ (Color online) Comparison between overlapping networks and
    networks with a hierarchical structure. We compare results from
    applying the three different methods (spectral decomposition, {\it
    k}-clique percolation, and modularity-landscape surveying) to (i)
    networks with a small overlap, and (ii) networks with a
    hierarchical structure. We show typical results for (i) a network
    with overlapping groups of 100 nodes, $z=16$, $r=0.1$, $p_2 =
    2\:p_1$, and $s = 0.1$ (top row---see Fig.~1 for details); (ii) a
    network with hierarchical structure such that at the top level
    there are three groups (two of 100 nodes and one of 200 nodes),
    and the largest group is comprised of two sub-groups of 100
    nodes. We construct the latter network such that $z=16$ and
    $r=0.1$, for nodes in the groups with flat organization. For edges
    involving nodes inside the large group, we link nodes $(i,j)$ with
    probability: (i) $p_1$ if they belong to the same sub-group, (ii)
    with $p_2 < p_1$ if they do not belong to the same sub-group, but
    belong to the same group at the top level; and, (iii) with $p_0$
    if they do not belong to the same group at the top level. Note
    that $p_0$ is the same for any pair of edges running across
    groups, and that $p_1$ and $p_2$ are selected such that $z=16$. We
    show results for the case $r_2= p_2/p_1=1/3$.  % Note that there
    are very few differences in the results for both networks. The
    spectral decomposition method finds four groups for both
    cases. The {\it k}-clique percolation method with $k=4$ subtly
    outlines the underlying group structure of the network shown by
    the adjacency matrix. Results for $k=5$ are not shown because
    there are no cliques of that size for sparse networks. Note that
    the signal detected by the {\it k}-clique percolation method has
    significantly decreased compared to that for the smaller networks
    in Fig.~(2).  In contrast, the signal detected by the
    modularity-landscape surveying method has not decreased. Note that
    results for hierarchical and overlapping networks are very
    similar.  }
  \label{f.comparison}
\end{figure*}

Like the spectral decomposition method, the modularity-landscape surveying 
method also indicates that there are three different groups for $s>0.2$.
Additionally, it is impossible to detect the overlap between groups from 
the collection of edges alone. The signal is only distinct for $s\le 0.1$,
and even then it is not clear whether one can distinguish between the
case in which two groups overlap and the case in which a group comprises 
two sub-groups (Fig.~\ref{f.comparison})~\cite{sales-pardo07}.

%%%%%%%%%%%%%%%%%%%%%%%%%%%%%%%%%%%%%%%%%%%%%%%%%
%Description of the conclusions
%%%%%%%%%%%%%%%%%%%%%%%%%%%%%%%%%%%%%%%%%%%%%%%%%
\section{Conclusions}
\label{sec:conclusions}
The ability to detect overlapping communities within real-world
networks would greatly enhance understanding of phenomena such as
synchronization~\cite{arenas06}. However, our analysis reveals that
the group detection methods in the literature are not entirely
equipped to handle such information. In some cases, these methods may
require tunable parameters, such as in the k-clique percolation
method~\cite{palla05} and the method of Gfeller et
al.~\cite{gfeller05}. The promising method of Nepusz et
al.~\cite{nepusz08} aims to obtain the global organization of a
network while determining which nodes act as ``bridges'' between
communities. This method captures some of the same information as the
modularity-landscape surveying method, but requires additional
centrality calculations to correct for the ``bridgeness''
score. Furthermore, even the recently proposed modularity-landscape
surveying method, which can detect small overlaps, is not able to
unambiguously differentiate overlapping groups from
hierarchically-organized groups.

%%%%%%%%%%%%%%%%%%%%%%%%%%%%%%%%%%%%%%%%%%%%%%%%%%%
%%%%%%%%%%%%%% ACKNOWLEDGMENTS %%%%%%%%%%%%%%%%%%%%
%%%%%%%%%%%%%%%%%%%%%%%%%%%%%%%%%%%%%%%%%%%%%%%%%%%
\section*{Acknowledgments}
We thank R. Guimer\`a, M.J. Stringer, and M.E.J. Newman for comments. 
L.A.N.A. gratefully acknowledges the support of the Keck
Foundation, an NIH/NIGMS K-25 award, and of NSF.

\begin{figure*}
\resizebox{2\columnwidth}{!}{
\includegraphics{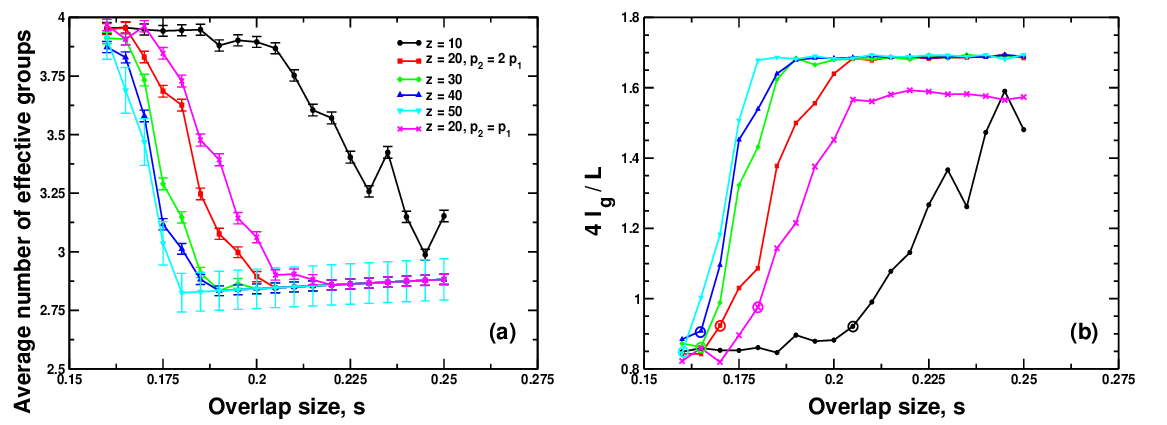}}
\caption{ (Color online) ({\bf a}), This graph shows the average
number of effective groups detected by spectral decomposition for a
range of overlap sizes and for different average degrees. The average
degrees investigated were $z=10,~20,~30,~40,~50$. Forty networks were
constructed for all cases except the $z=50$ case, which only had ten
networks for every overlap size. Each network consists of four groups,
two of which share nodes. As the network becomes more dense with
increasing degree, the transition from the detection of approximately
four groups to three groups occurs abruptly at an overlap size of
$s=0.16$.
%}
%\label{f.degree_effects}
%\end{figure}
%%%%%%%%%%%%%%%%%%%%%%%%%%%%%%%%%%%%%%%%%%%%%%%%%%%%%%%%%%%%%%%%%%%%%%%%%
%\newpage
%%%%%%%%%%%%%%%%%% Resolution limit Figure %%%%%%%%%%%%%%%%%%%%%%%%%%%%
%%%%%%%%%%%%%%%%%%%%% p2 = 2p1  %%%%%%%%%%%%%%%%%%%%%%%%%%%%%%
%\begin{figure}
%\resizebox{\columnwidth}{!}{
%\includegraphics{figureS3}}
%\caption{
({\bf b}), This graph shows the resolution of limit detectability. In
the paper by Fortunato and Barth\'{e}lemy, they determined that for a
module to be detectable via spectral decomposition, the number of
links within the module, $l_g$, should be less than $\frac{L}{4}$
where $L$ is the total number of links in the network. We tested these
conditions for several networks with overlapping groups by
construction. The size of the overlap varied from $s=0.16$ to
$s=0.25$. Finally, forty networks were generated for each set of
parameters except for $z=50$, in which ten networks were generated. In
this case, we plot $4 \frac{l_g}{L}$ for each overlap size and degree,
and see that the transition of detectability occurs over the same
range as is indicated in the average number of effective
groups. The transition sites are circled. }
\label{f.resolution_limits}
\end{figure*}

%%%%%%%%%%%%%%%%%%%%%%%%%%%%%%%%%%%%%%%%%%%%%%%%%%%
%%%%%%%%%%%%%%%%%%% APPENDIX %%%%%%%%%%%%%%%%%%%%%%
%%%%%%%%%%%%%%%%%%%%%%%%%%%%%%%%%%%%%%%%%%%%%%%%%%%
\appendix{{\bf Appendix A}

\noindent
To characterize how the degree and the size of the overlap affect
group detection, we generate forty networks composed of four groups of
100 nodes each for every set of conditions tested. We tested the
degree of a non-overlapping node at values of $z=10, 20, 30, 40, 50$,
and we tested the size of the overlap, $s$, for $s=0.16$ through
$s=0.25$. Each of these networks was generated with $p_2=2p_1$, except
for the case $z=20$, for which we also studied networks with
$p_2=p_1$.

Since the spectral decomposition method is very fast and is considered
the gold standard for group detection, we applied this method on
each of the networks generated (Fig.~\ref{f.resolution_limits}a). We
expected the spectral decomposition method to detect four groups for
low overlap sizes for every degree tested, and that it would detect
three groups at higher overlap sizes or at higher node degree. We
expected also that the higher the degree, the faster the transition
between four detectable groups and three detectable groups. Since some
of the groups reported by the method were very small, we calculated
the effective number of groups, $N_{\rm effective}$.

\begin{equation}
N_{\rm effective} = \frac{1}{\sum_{\mathcal{G}} (\frac{S_i}{N})^2}
\end{equation}
where $\mathcal{G}$ is the list of groups in the partition of the network 
returned by the method, $S_i$ is the number of nodes in a group within $\mathcal{G}$, 
and $N$ is the total number of nodes in the network.
For the networks with an average degree of $z=20$ and $p_2 = p_1$, 
we find a transition from detecting four groups to detecting three groups 
for $s \approx 0.18$, while for $z=50$, the transition occurs by $s=0.16$.

However, we wanted to further investigate the detectability for the
degrees and overlap sizes chosen. Specifically, we examine the
detection resolution limits as outlined by Fortunato and
Barth\'{e}lemy in \cite{fortunato07}. In their paper, Fortunato and
Barth\'{e}lemy indicate that in order for a group to be unambiguously
detectable by spectral decomposition, it must meet two criteria: ({\em
i}) the number of links within group $g$, or $l_g$, should be less
than the total number of links $L$ divided by four, or $l_g <
\frac{L}{4}$; and ({\em ii}) the ratio of links leaving the module,
$l_g^{out}$, to links within the module should be less than two, or $a
= \frac{l_g^{out}}{l_g} < 2$. All of the detected groups for each of
the networks satisfied the second
condition. Figure~\ref{f.resolution_limits} shows $4*\frac{l_g}{L}$
versus $s$ for each degree. In this figure, a value less than one
indicates that the resolution detection limit is satisfied.  Comparing
Figures~\ref{f.resolution_limits}a and b, we see that as
soon as the resolution detection limit is violated, the number of
detected groups decreases, so our results are consistent with those of
Fortunato and Barth\'{e}lemy~\cite{fortunato07}.
}
%%%%%%%%%%%%%%%%%%%%%%%%%%%%%%%%%%%%%%%%%%%%%%%%%%%
%%%%%%%%%%%%%%%%%% FIGURES %%%%%%%%%%%%%%%%%%%%%%%%
%%%%%%%%%%%%%%%%%%%%%%%%%%%%%%%%%%%%%%%%%%%%%%%%%%%

\clearpage
%%%%%%%%%%%%%%%%% Overlapping Model Schematics %%%%%%%%%%%%%%%%%%%%%%%%%%%%
%%%%%%%%%%%%%%%%%%%%%%%%%%%%%%%%%%%%%%%%%%%%%%%%%%%%%%%%%%%%%%%%%%%%%%%%

\newpage
%%%%%%%%%%%%%%% Coclassification Matrices for Both Models %%%%%%%%%%%%%%%
%%%%%%%%%%%%%%%%%%%%%%%%%%%%%%%%%%%%%%%%%%%%%%%%%%%%%%%%%%%%%%%%%%%%%%%%

\newpage
%%%%%%%%%%%%%%%%% Overlap Mutual Info Figure %%%%%%%%%%%%%%%%%%%%%%%%%%%%
%%%%%%%%%%%%%%%%%%% p2 = 1p1: large overlap %%%%%%%%%%%%%%%%%%%%%%%%%%%%%
%%%%%%%%%%%%%%%%%%%%%%%%%%%%%%%%%%%%%%%%%%%%%%%%%%%%%%%%%%%%%%%%%%%%%%%%

\newpage
%%%%%%%%%%%%%%%%%% Resolution limit Figure %%%%%%%%%%%%%%%%%%%%%%%%%%%%
%%%%%%%%%%%%%%%%%%%%% p2 = 2p1  %%%%%%%%%%%%%%%%%%%%%%%%%%%%%%
%%%%%%%%%%%%%%%%%%%%%%%%%%%%%%%%%%%%%%%%%%%%%%%%%%%%%%%%%%%%%%%%%%%%%%%%%

%%%%%%%%%%%%%%% Hierarchical Comp Figure %%%%%%%%%%%%%%%%%%%%%
%%%%%%%%%%%%%%%%%%%%% p2 = 2p1  %%%%%%%%%%%%%%%%%%%%%%%%%%%%%%
%%%%%%%%%%%%%%%%%%%%%%%%%%%%%%%%%%%%%%%%%%%%%%%%%%%%%%%%%%%%%%%%%%%%%%%%%

\newpage
%%%%%%%%%%%%%%%%%% Degree Effects Figure %%%%%%%%%%%%%%%%%%%%%%%%%%%%
%%%%%%%%%%%%%%%%%%%%% p2 = 2p1  %%%%%%%%%%%%%%%%%%%%%%%%%%%%%%
%%%%%%%%%%%%%%%%%%%%%%%%%%%%%%%%%%%%%%%%%%%%%%%%%%%%%%%%%%%%%%%%%%%%%%%%%

%%%%%%%%%%%%%%%%%%%%%%%%%%%%%%%%%%%%%%%%%%%%%%%%%%%
%%%%%%%%%%%%%% REFERENCES %%%%%%%%%%%%%%%%%%%%%%%%%
%%%%%%%%%%%%%%%%%%%%%%%%%%%%%%%%%%%%%%%%%%%%%%%%%%%

\bibliographystyle{epj}
\bibliography{References/ref-database,ref-specific}

\begin{thebibliography}{27}

\bibitem{newman04b}
M.E.J. Newman, M.~Girvan, Phys. Rev. E \textbf{69}(2), art. no. 026113 (2004)

\bibitem{guimera05}
R.~Guimer\`a, L.A.N. Amaral, J. Stat. Mech.: Theor. Exp. p. art. no. P02001
  (2005)

\bibitem{watts02}
D.J. Watts, P.S. Dodds, M.E.J. Newman, Science \textbf{296}, 1302 (2002)

\bibitem{donetti04}
L.~Donetti, M.A. \protect{Mu\~{n}oz}, J. Stat. Mech.: Theor. Exp. p. art. no.
  P10012 (2004)

\bibitem{guimera04}
R.~Guimer\`a, M.~Sales-Pardo, L.A.N. Amaral, Phys. Rev. E \textbf{70}, art. no.
  025101 (2004)

\bibitem{reichardt04}
J.~Reichardt, S.~Bornholdt, Phys. Rev. Lett. \textbf{93}, art. no. 218701
  (2004)

\bibitem{duch05}
J.~Duch, A.~Arenas, Phys. Rev. E \textbf{72}, art. no. 027104 (2005)

\bibitem{palla05}
G.~Palla, I.~Der\'enyi, I.~Farkas, T.~Vicsek, Nature \textbf{435}(7043), 814
  (2005)

\bibitem{baumes05}
J.~Baumes, M.~Goldberg, M.~Magdon-Ismail, \emph{Efficient Identification of
  Overlapping Communities}, in \emph{Intelligence and Security Informatics:
  IEEE International Conference on Intelligence and Security Informatics, ISI
  2005}, edited by P.~Kantor, G.~Muresan, F.~Roberts, D.~Zeng, F.Y. Wang,
  H.~Chen, R.~Merkle, IEEE (Springer, Berlin, 2005), Vol. 3495 of \emph{Lecture
  Notes in Computer Science}, pp. 27--36

\bibitem{gfeller05}
D.~Gfeller, J.C. Chappelier, P.D.L. Rios, Phys. Rev. E \textbf{72}, 056135
  (2005)

\bibitem{nepusz08}
T.~Nepusz, A.~Petr\'{o}czi, L.~N\'{e}gyessy, F.~Bazs\'{o}, Phys. Rev. E
  \textbf{77}, 016107 (2008)

\bibitem{pereira-leal04}
J.B. Pereira-Leal, A.J. Enright, C.A. Ouzounis, Proteins \textbf{54}(1), 49
  (2004)

\bibitem{sales-pardo07}
M.~Sales-Pardo, R.~Guimer\`a, A.A. Moreira, L.A.N. Amaral, Proc. Natl. Acad.
  Sci. U. S. A. \textbf{104}, 15224 (2007)

\bibitem{clauset08}
A.~Clauset, C.~Moore, M.E.J. Newman, Nature \textbf{453}(7191), 98 (2008)

\bibitem{guimera05a}
R.~Guimer\`a, L.A.N. Amaral, Nature \textbf{433}, 895 (2005)

\bibitem{newman06}
M.E.J. Newman, Proc. Natl. Acad. Sci. USA \textbf{103}(23), 8577 (2006)

\bibitem{newman03h}
M.E.J. Newman, Phys. Rev. E \textbf{68}, 026121 (2003)

\bibitem{guimera05b}
R.~Guimer\`a, S.~Mossa, A.~Turtschi, L.A.N. Amaral, Proc. Natl. Acad. Sci. USA
  \textbf{102}(22), 7794 (2005)

\bibitem{hastings06}
M.B. Hastings, Phys. Rev. E \textbf{74}(3 Pt 2), art. no. 035102 (2006)

\bibitem{newman06b}
M.E.J. Newman, Phys. Rev. E \textbf{74}(3 Pt 2), art. no. 036104 (2006)

\bibitem{fortunato07}
S.~Fortunato, M.~Barth\'elemy, Proc. Natl. Acad. Sci. USA \textbf{104}(1), 36
  (2007)

\bibitem{kumpula07}
J.M. Kumpula, J.~Saramaki, K.~Kaski, J.~Kertesz, Eur. Phys. J. B \textbf{56},
  41 (2007)

\bibitem{danon05}
L.~Danon, A.~D\'{\i}az-Guilera, J.~Duch, A.~Arenas, J. Stat. Mech.: Theor. Exp.
  p. art. no. P09008 (2005)

\bibitem{stillinger82}
F.H. Stillinger, T.A. Weber, Phys. Rev. A \textbf{25}, 978 (1982)

\bibitem{stillinger84}
F.H. Stillinger, T.A. Weber, Science \textbf{225}, 983 (1984)

\bibitem{stillinger85}
Stillinger, Weber, Phys. Rev. B \textbf{31}(8), 5262 (1985)

\bibitem{arenas06}
A.~Arenas, A.~D\'iaz-Guilera, C.J. P\'erez-Vicente, Phys. Rev. Lett.
  \textbf{96}, art. no. 114102 (2006)

\end{thebibliography}

\end{document}